# Outage Regions and Optimal Power Allocation for Wireless Relay Networks

Tobias Renk
Institut für Nachrichtentechnik
Universität Karlsruhe (TH)
Email: renk@int.uni-karlsruhe.de

Holger Jaekel
Institut für Nachrichtentechnik
Universität Karlsruhe (TH)
Email: jaekel@int.uni-karlsruhe.de

Friedrich K. Jondral
Institut für Nachrichtentechnik
Universität Karlsruhe (TH)
Email: fj@int.uni-karlsruhe.de

*Abstract*—We study outage regions for energy-constrained multi-hop and adaptive multi-route networks with an arbitrary number of relay nodes. Optimal power allocation strategies in the sense that outage probability is minimized are derived depending on the distances between the transmit nodes. We further investigate the rate gain of adaptive multi-route and multi-hop over direct transmission. It is shown that a combined strategy of direct transmission and adaptive multi-route outperforms multi-hop for all values of rate $R$. It can be stated that cooperation strategies are beneficial for low-rate systems where the main goal is a very low outage probability of the network. As the rate is increased, direct transmission becomes more and more attractive.

## I. INTRODUCTION

Cooperation has been introduced as a proper means to mitigate fading effects on wireless channels which lead to severe fluctuations of the received signal's amplitude. The basic idea is that several transmit nodes pool their resources in order to create a "virtual" antenna array and exploit spatial diversity at the destination (see [1], [2], [3], [4], [5] and the references therein). The research topics vary from coding strategies over relay selection algorithms [6] to combining receivers – just to name a few. In this paper we deal with outage regions, optimal power allocation, and rate gain of cooperative transmission schemes over direct transmission. Particularly, we consider multi-hop networks, where there is no direct link between source and destination, and adaptive multi-route networks, where direct communication between source and destination is possible and the relay only aids communication if it has been able to decode the source message reliably.

In general, a combination of the mentioned cooperation strategies is best suited. The reason for this is that mostly one either increases the data rate and thus reduces reliability or increases reliability at the cost of a reduced coverage area. This issue is solved in cellular networks, for instance, by shrinking the cell size and installing additional base stations which means additional costs for antenna space at the base stations and for the wired backhaul network. An alternative solution is the insertion of (fixed) relays that only aid communication from a base station to a mobile station and vice versa. This kind of network is often referred to as multi-hop cellular network [7].

*Related Work and Main Contributions:* The idea of outage regions for cooperative networks has been first introduced, to the authors' best knowledge, in [8]. There, the authors consider essentially the same cooperation strategies. However, they only deal with networks that consist of one source, one relay, and one destination node. In [9] outage regions are defined such that all values below the outage curve need more power than available to guarantee no outage. The authors derive a power control policy for the "classical" three-node relay channel for amplify-and-forward. Another publication that deals with optimal power allocation in the sense of outage minimization is [10]. Here, the authors derive expressions on optimal power allocation for high values of signal-to-noise ratio (SNR). In contrast to the mentioned papers, we give expressions on outage regions for networks with an arbitrary number of relay nodes. Moreover, we get an optimal power allocation strategy $\beta_{\text{opt}}$ with respect to minimization of outage probability depending on the distance between nodes. Information on the distance between the nodes can be gathered by estimation of path-losses via training sequences, where all nodes transmit with a predefined power during an initialization phase. Additionally, the metric rate gain $r_\epsilon$, which describes the rate difference between two systems for the same outage probability $\epsilon$, is introduced and studied. Interestingly, we demonstrate that a combined strategy of adaptive multi-route and direct transmission outperforms multi-hop in all cases of rate $R$. Hence, multi-hop is only suitable for range extension but not for an increase in data rate.

## II. SYSTEM MODEL

We consider networks that consist of one source node S, $K$ relay nodes $\text{R}_k$, $k = 1, \ldots, K$, and one destination node D. The set of all relay nodes is given by $\mathcal{R}$ and the set of all relay nodes that have been able to decode the source message reliably is given by $\mathcal{K} \subseteq \mathcal{R}$. The communication channel between two terminals $i$ and $j$ is modeled as a flat fading Rayleigh channel with additive white Gaussian noise components which means that fading influences remain constant over one transmission period $T$. The fading coefficients $h_{ij}$ are zero-mean, independent, circularly-symmetric complex Gaussian random variables, where the real and imaginary parts are uncorrelated with variance $\sigma_{ij}/2$ each. Hence, the magnitudes $|h_{ij}|$ follow a Rayleigh distribution and the channel powers $|h_{ij}|^2$ possess an exponential distribution with mean value $\mathbb{E}\{|h_{ij}|^2\} = \sigma_{ij}^2$. The phases $\arg(h_{ij})$ are uniformly distributed on $[0, 2\pi)$. We employ a common path-loss model, $\sigma_{ij}^2 \propto d_{ij}^{-\alpha}$, where $d_{ij}$ denotes the distance between node $i$ and $j$ and $\alpha \in [3, 5]$ is the path-loss exponent for shadowed urban cellular radio scenarios. Moreover, decode-and-forward is employed at the relay nodes, which means that each relay itself sends a "refreshed" version of its receive signal. On each path additive white Gaussian noise (AWGN) with one-sided power spectral density $N_0$ is added and we define $\mathsf{SNR} := P/N_0$, where $P$ is the maximal transmit power of a node. For fair comparison we want the end-to-end target rate $R$ to be the same for all transmission schemes. Mostly, an average power constraint per transmit node is considered. However, in the context of ad-hoc networks, an energy constraint makes much more sense [11]. Therefore, the total energy $E$ over one block of duration $T$ becomes

$$E = \sum_{k=0}^{K} P_k T_k, \qquad (1)$$

where $P_k$ is the transmit power and $T_k$ is the transmission time of node $k$, respectively. In the following, we use the subscript 0 for the source node, the subscripts 1 to $K$ for the relay nodes, and $K+1$ for the destination node. We allow each node to transmit over orthogonal time slots of duration $T_k = T/(K+1)$. In order to achieve an energy constraint, transmit power of node $k$ is then given by $P_k = (K+1)\beta_k P$, where $\beta_k$ describes the power allocation fraction. Hence,

$$E = \sum_{k=0}^{K} \beta_k P T. \qquad (2)$$

A power allocation strategy is given by the vector $\boldsymbol{\beta} = (\beta_0, \ldots, \beta_K)^T$, where $^T$ denotes transposition of a vector. Clearly, the energy constraint is fulfilled for $|\boldsymbol{\beta}|_1 = \sum_{k=0}^{K} \beta_k = 1$, where $|\cdot|_1$ is the $L_1$-norm of a vector.

Two transmission schemes are considered in the paper. First, multi-hop networks, where a transmit node can only send its message to the adjacently located node. Without loss of generality, we assume that the relays occur in an ordered manner. Hence, for multi-hop networks, the source S sends its message to the first relay $R_1$. The first relay sends its message to $R_2$ and so on. Second, adaptive multi-route networks. Here, we have a direct link from source to destination. Additionally, each relay, that has been able to decode the source message aids in communication. As stated before, one transmission block is divided into $K+1$ subblocks of equal lengths. Those blocks that originally belonged to relays that could not decode the source message are then allocated to the source again. This is shown for the case of 2 relays in Fig. 1. Subplot (a) shows the case where both relays have not been able to decode. Therefore, each subblock is used by the source which transmits in each subblock with a rate of $3R$. In subplot (b) and (c), $R_1$ and $R_2$, respectively, could not decode. The corresponding subblock is occupied by the source. In subplot (d), each relay

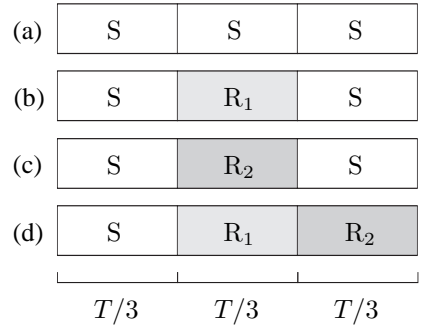

Fig. 1. Transmission scenarios depending on the ability of the relays $R_1$ and $R_2$ to decode. If a relay cannot decode, the source S sends in the corresponding subblock again. (a) Both relays cannot decode, (b) relay $R_2$ cannot decode, (c) relay $R_1$ cannot decode, (d) both relays can decode.

has been able to decode and, hence, the source only occupies the first subblock.

### III. OUTAGE REGION

#### A. Definition

Let $\mathbf{h}$ denote the vector that contains all channel gains of the network. Then we have the following definition.

*Definition 1:* The outage region $\mathcal{O}_K$ of a network with $K$ relays is the set of channel gains $\mathbf{h}$ for which the instantaneous channel capacity $\mathcal{C}(\mathbf{h})$ is not able to support a desired target rate $R$:

$$\mathcal{O}_K := \{\mathbf{h} : \mathcal{C}(\mathbf{h}) < R\} \qquad (3)$$

The outage region can be interpreted as a volume in a $2K+1$-dimensional space, where channel conditions are such that the available transmit power is not high enough to invert the channel influence and ensure reliable communications. However, it must be stated that the pure shape and size of the volume are not enough to draw conclusions about the efficiency of a particular relay strategy; the statistical characteristics of the channel gains, i.e., their distribution, also have to be taken into account. For multi-hop networks, we have $\mathbf{h} = (h_{01}, h_{12} \ldots, h_{K(K+1)})^T$, and for multi-route networks, we have $\mathbf{h} = (h_{01}, \ldots, h_{0(K+1)}, h_{1(K+1)}, \ldots, h_{K(K+1)})^T$.

#### B. Multi-Hop Networks

The instantaneous channel capacity in bit/channel use (bit/s/Hz) of multi-hop networks with $K$ relays is given by

$$\mathcal{C}_{\mathrm{MH}} = \min\{\mathcal{C}_{01}, \mathcal{C}_{12}, \ldots, \mathcal{C}_{(K-1)K}, \mathcal{C}_{K(K+1)}\}, \qquad (4)$$

where

$$\mathcal{C}_{k(k+1)} = C_K(\beta_k |h_{k(k+1)}|^2 \mathsf{SNR}) \qquad (5)$$

and $C_K(x) = (K+1)^{-1} \log_2(1 + (K+1)x)$. This shows that for multi-hop networks the achievable channel capacity is limited by the weakest channel between the nodes (cf. [12]). With Definition 1, the outage region becomes

$$\mathcal{O}_{\mathrm{MH}}(\mathbf{h}, \boldsymbol{\beta}) = \{\mathbf{h} : \min\{\beta_k |h_{k(k+1)}|^2\} < \gamma_K\}, \qquad (6)$$

where we used

$$\gamma_K = \frac{2^{(K+1)R} - 1}{(K+1)\mathsf{SNR}}$$

and omitted the dependence on $R$ and SNR in the description for reasons of presentation. Nonetheless, we used $\mathcal{O}_{\text{MH}}(\mathbf{h}, \boldsymbol{\beta})$ to clearly point out the dependence of the outage region on the power allocation. We will later use this dependency for optimization tasks.

### C. Adaptive Multi-Route Networks

The instantaneous channel capacity of an adaptive multi-route network is given by

$$\mathcal{C}_{\text{AMR}} = \begin{cases} C_K(|h_{0(K+1)}|^2 \text{SNR}) & : \mathcal{A} \\ C_K((\beta_0 |h_{0(K+1)}|^2 + \sum_{k \in \mathcal{K}} \beta_k |h_{k(K+1)}|^2)\text{SNR}) & : \mathcal{B} \\ C_K((\beta_0 |h_{0(K+1)}|^2 + \sum_{k \in \mathcal{R}} \beta_k |h_{k(K+1)}|^2)\text{SNR}) & : \bar{\mathcal{A}} \end{cases}$$

where the event $\mathcal{A}$ describes that all relays have not been able to decode the source message and hence the source allocates all $K+1$ subblocks,

$$\mathcal{A} = \{h_{0k} : \beta_0 |h_{0k}|^2 < \gamma_K \, \forall k \in \mathcal{R}\},$$

the event $\mathcal{B}$ describes that some relays have been able to decode ($k \in \mathcal{K}$) and the rest has not ($l \in \mathcal{L} := \mathcal{R} \setminus \mathcal{K}$),

$$\begin{aligned} \mathcal{B} = & \{(h_{0k}, h_{0l}) : \beta_0 |h_{0k}|^2 \geq \gamma_K \, \forall k \in \mathcal{K} \\ & \text{ and } \beta_0 |h_{0l}|^2 < \gamma_K \, \forall l \in \mathcal{L}\}, \end{aligned}$$

and $\bar{\mathcal{A}}$ is the complement of $\mathcal{A}$, i.e., all relays have been able to decode the source message and send subsequently to the destination. Considering $\mathcal{C}_{\text{AMR}}$, the outage region of adaptive multi-routing for an arbitrary number of relay nodes is given by[1]

$$\begin{aligned} \mathcal{O}_{\text{AMR}}(\mathbf{h}, \boldsymbol{\beta}) = & \{\mathbf{h} : [\mathcal{A} \cap (|h_{0(K+1)}|^2 < \gamma_K)] \\ & \cup [\mathcal{B} \cap (\beta_0 |h_{0(K+1)}|^2 + \sum_{k \in \mathcal{K}} \beta_k |h_{k(K+1)}|^2 < \gamma_K)] \quad (7) \\ & \cup [\bar{\mathcal{A}} \cap (\beta_0 |h_{0(K+1)}|^2 + \sum_{k \in \mathcal{R}} \beta_k |h_{k(K+1)}|^2 < \gamma_K)]\}. \end{aligned}$$

We see that the outage regions for multi-hop and adaptive multi-route depend on $\boldsymbol{\beta}$ and, hence, can be optimized with respect to power allocation. This means that there exists an allocation strategy $\boldsymbol{\beta}_{\text{opt}}$ that minimizes the volume of the outage region and, thus, also minimizes the outage probability. This is shown in more detail in the next section.

## IV. OPTIMAL POWER ALLOCATION

### A. Definitions

In the following paragraphs we define outage probability with respect to the outage region given in Definition 1 and derive an expression on the optimal power allocation strategy.

*Definition 2:* Outage probability $p_{\text{out}}$ is the probability that the instantaneous channel capacity $\mathcal{C}(\mathbf{h})$ cannot support a required target rate $R$:

$$p_{\text{out}} = \Pr(\mathcal{C}(\mathbf{h}) < R) = \int_{\mathcal{O}_K} f_{\textsf{SNR}}(\textsf{SNR}) \, d\textsf{SNR}, \quad (8)$$

[1]We stress that we used a rather sloppy notation for the outage region of adaptive multi-route here for the sake of compactness.

where $\textsf{SNR} \in \mathbb{R}^{2K+1}$, $\textsf{SNR}$ represents a vector containing all instantaneous SNR values, and $f_{\textsf{SNR}}(\textsf{SNR})$ is the joint probability density function.

*Definition 3:* The optimal power allocation strategy $\boldsymbol{\beta}_{\text{opt}}$ minimizes the outage probability under a given network energy constraint. Hence,

$$\boldsymbol{\beta}_{\text{opt}} := \arg \min_{\substack{\boldsymbol{\beta} \\ |\boldsymbol{\beta}|_1 = 1}} p_{\text{out}}(\boldsymbol{\beta}). \quad (9)$$

Clearly, $\boldsymbol{\beta}_{\text{opt}}$ depends on the average channel gains, which is equivalent to the distances between transmit and receive nodes in our system model, and requires channel state information at the transmit nodes.

### B. Distance-dependent Power Allocation

Distance-dependent power allocation is a well-known scheme [13], [14]. Generally, there are two possibilities how a mobile node can gain information about its location. The first possibility is based on the global positioning system (GPS). Here, the relay has information about its own location, but no information about the location of other nodes. Since power allocation in our sense does not only depend on the location of a single node, but rather on the distances between the nodes, this scheme is not suitable for our purposes. Of course, by introducing overhead and putting more effort into this scheme, the performance can be increased enormously. A more practical scheme, however, is the estimation of path-losses via training sequences. Here, the mobile nodes transmit in an initialization phase with a predefined transmit power and send an a-priori known bit sequence. Surrounding nodes are now able to estimate the distance to this node. After the initialization phase, a node knows its distance to the destination and to other nodes and can adjust its transmit power by selecting the corresponding value from a look-up table.

## V. EXAMPLES

In this section we give performance examples of the investigated strategies for the case of one relay. Thus, we have $\boldsymbol{\beta} = (\beta_0, \beta_1)^T$. In order to meet the energy constraint, we must set $\beta_0 = 1 - \beta_1$. For the sake of presentation, we use $\beta_0 = \beta$ and $\beta_1 = 1 - \beta$. The relay is placed on a straight line between source and destination, and the source-to-destination distance is normalized to 1. Consequently, we get $d_{\text{rd}} = 1 - d_{\text{sr}}$, where $d_{\text{rd}}$ is the relay-to-destination distance and $d_{\text{sr}}$ is the source-to-relay distance. Some of the results have also been reported in [10]. However, especially for AMR, the optimal power fraction has only been given for large values of SNR.

The optimal power fraction can be derived by minimization of the outage probability. For multi-hop, we get

$$\beta_{\text{opt}}^{(\text{MH})} = \frac{1}{1 + \sqrt{\left(\frac{d_{\text{rd}}}{d_{\text{sr}}}\right)^{\alpha}}}, \quad (10)$$

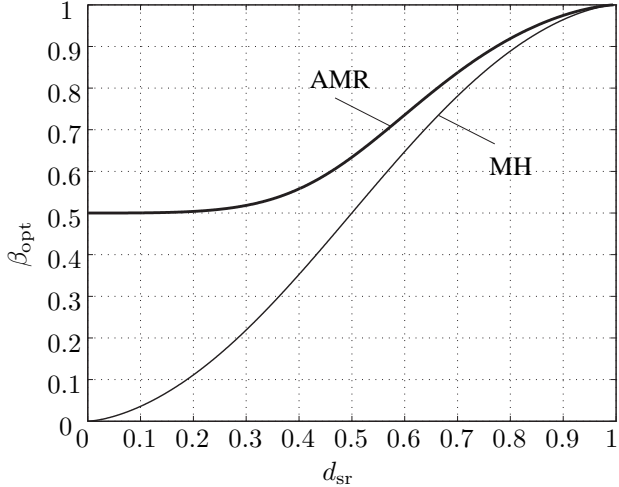

Fig. 2. Optimal power allocation strategy versus relay location for multi-hop (MH) and adaptive multi-route (AMR) networks for $\alpha = 3$.

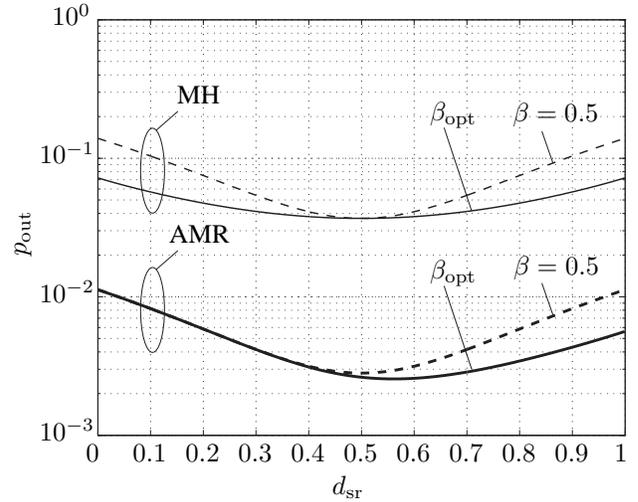

Fig. 3. Outage probability for $\beta = 0.5$ and $\beta_{\text{opt}}$ versus relay location for multi-hop (MH) and adaptive multi-route (AMR) networks. Parameters are $\alpha = 3$, $R = 2$ bit/s/Hz, and $\mathsf{SNR} = 10$ dB.

which ranges from 0 to 1 dependent on the relay location. It can easily be seen that

$$\lim_{d_{\text{sr}} \to 0} \beta_{\text{opt}}^{(\text{MH})} = 0 \quad \text{and} \quad \lim_{d_{\text{sr}} \to 1} \beta_{\text{opt}}^{(\text{MH})} = 1. \qquad (11)$$

If the relay is located close to the source, the channel between source and relay will be good and therefore little transmit power has to be allocated to the source in order to have reliable decoding at the relay. However, the distance from relay to destination is comparably large, that is why more transmit power has to be allocated to the relay. If the relay is located half-way between source and destination, the optimal power allocation strategy is $\beta = 0.5$, which means that both source and relay transmit with the same power. If the relay is located close to the destination, almost all transmit power is allocated to the source.

For adaptive multi-route, the optimal power allocation $\beta_{\text{opt}}^{(\text{AMR})}$ becomes

$$\beta_{\text{opt}}^{(\text{AMR})} = \frac{1}{2}\left[2 + \left(\frac{d_{\text{rd}}}{d_{\text{sr}}}\right)^{\alpha} - \sqrt{\left(\frac{d_{\text{rd}}}{d_{\text{sr}}}\right)^{2\alpha} + 2\left(\frac{d_{\text{rd}}}{d_{\text{sr}}}\right)^{\alpha}}\right]. \qquad (12)$$

It can easily be seen that $\lim_{d_{\text{sr}} \to 1} \beta_{\text{opt}}^{(\text{AMR})} = 1$. For $d_{\text{sr}} \to 0$, $\beta_{\text{opt}}^{(\text{AMR})}$ becomes 0.5. In order to show that, we first define $x := (d_{\text{rd}}/d_{\text{sr}})^{\alpha}$. We then have to show that

$$\lim_{x \to \infty} x - \sqrt{x^2 + 2x} = -1. \qquad (13)$$

Especially, we have

$$\lim_{x \to \infty} x - \sqrt{x^2 + 2x} = \lim_{x \to \infty} x \left(1 - \sqrt{1 + \frac{2}{x}}\right)$$
$$= \lim_{x \to \infty} x \left(1 - \left(1 + \frac{1}{x} + R_2(x)\right)\right) = -1,$$

where $R_2(x)$ is the remainder term of a Taylor series development and goes to zero for large values of $x$. The interpretation is as follows. When the relay is located close to the source, both nodes face more or less the same channel and a proper allocation strategy is that both nodes send with equal power. The more the relay moves to the destination, the more power is given to the source since it faces a more severe channel then. Fig. 2 illustrates the optimal power allocation strategy versus the relay location for multi-hop and adaptive multi-route. In Fig. 3 outage probability versus relay location is shown. As expected, outage probability for multi-hop is symmetric to $d_{\text{sr}} = 0.5$. When the relay is placed half-way between source and destination $\beta_{\text{opt}} = 0.5$. This is not the case for adaptive multi-route. In the range from $d_{\text{sr}} = 0$ to $d_{\text{sr}} = 0.5$, equal power allocation is almost optimal which can be seen by the fact that within that region both outage probabilities are nearly the same. The above mentioned investigations are only true for a fixed rate $R$. However, with the ever-increasing demand for higher data rates, it is indispensable to compare cooperation strategies with respect to their outage behavior if the rate increases. For that purpose we define a novel metric named rate gain.

*Definition 4:* The rate gain $r_\epsilon(A, B)$ in bit/s/Hz of system $A$ over system $B$ for an outage probability $\epsilon$ is defined as

$$r_\epsilon(A, B) := R_A(\epsilon) - R_B(\epsilon). \qquad (14)$$

For $r_\epsilon < 0$ system $B$ outperforms system $A$, i.e., system $B$ can transmit with a higher rate than system $A$ and achieves the same outage probability. For $r_\epsilon = 0$ both systems show the same performance and for $r_\epsilon > 0$ system $A$ outperforms system $B$.

Rate gain is related to the SNR gain exponent $\zeta_\infty$ mentioned in [15]. There, it is shown that the slope of SNR gain for high values of rate $R$ only depends on the number of transmission phases. Fig. 4 shows outage probability versus rate. It can be seen that adaptive multi-route outperforms multi-hop to a rate of $R \approx 4.5$ bit/s/Hz. If the rate is increased more, both strategies show the same behavior. Direct transmission performs worst of all three strategies up to a rate of $R \approx 1.5$ bit/s/Hz. From that rate on, multi-hop is outperformed by direct transmission and adaptive multi-route. This clearly shows that multi-hop is not a suitable method for rate increase. Instead, multi-hop should rather be used for range extension, which is, for instance, the case for multi-hop cellular networks [7], [16]. Adaptive multi-route achieves a lower outage probability compared to direct transmission for rates up to approximately 3.3 bit/s/Hz. From that rate on, both cooperation strategies cannot perform as well as direct transmission with respect to outage probability. The reason for this lies in the model that we applied. In order to have a fair comparison, we have the same overall network energy in all systems and we ensure that the amount of information sent through all systems is the same, i.e., the number of transmitted bits. To ensure this, all transmit nodes in cooperative systems with one source node and $K$ relay nodes have to transmit with rate $R' = (K+1)R$. Summarizing, we see that a combined strategy of direct transmission and adaptive multi-route outperforms multi-hop for all values of rate $R$. In Tab. I rate gain $r_\epsilon$ for different values of outage probability $\epsilon$ is shown confirming that for high-rate systems direct transmission is beneficial to multi-hop and adaptive multi-route.

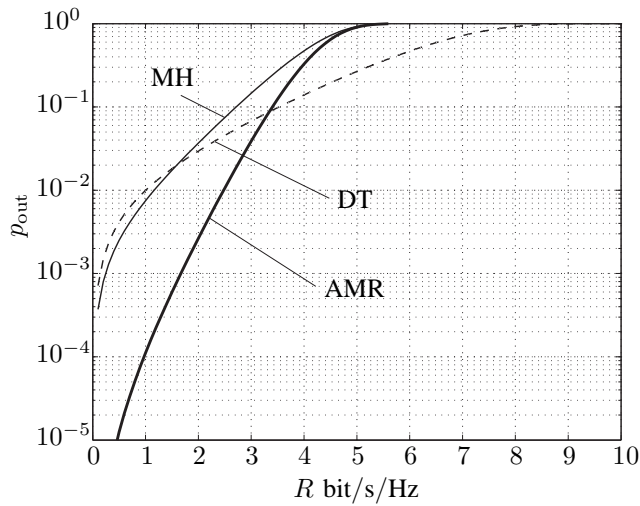

Fig. 4. Outage probability versus rate $R$ in bit/s/Hz for direct transmission (DT), multi-hop (MH) and adaptive multi-route (AMR) networks with optimal power allocation $\beta_{\text{opt}}$. Parameters are $\mathsf{SNR} = 30$ dB, $d_{\text{sr}} = 0.5$, and $\alpha = 3$.

TABLE I
RATE GAIN OF ADAPTIVE MULTI-ROUTE (AMR) AND MULTI-HOP (MH) OVER DIRECT TRANSMISSION (DT) FOR DIFFERENT VALUES OF OUTAGE PROBABILITY $\epsilon$. PARAMETERS CF. FIG. 4.

| $\epsilon$ | $r_\epsilon$ [bit/s/Hz] (AMR-DT) | $r_\epsilon$ [bit/s/Hz] (MH-DT) |
|---|---|---|
| $10^{-1}$ | $-0.1$ | $-0.9$ |
| $10^{-2}$ | $1.3$ | $-0.2$ |
| $10^{-3}$ | $1.7$ | $0.1$ |

## VI. CONCLUSIONS

In this paper we investigated outage regions for energy constraint multi-hop and adaptive multi-route networks with an arbitrary number of relay nodes. We then derived expressions for a distance-dependent optimal power allocation which minimizes outage region for cooperative networks with one relay node. This minimization is equivalent to a minimization of outage probability. We stress that knowledge about the distances between nodes can be gathered by estimation of pathlosses via training sequences. We further studied the rate gain of adaptive multi-route and multi-hop over direct transmission with the result that a combined strategy of direct transmission and adaptive multi-route outperforms multi-hop for all values of rate $R$. Generally, one can state that cooperation strategies are beneficial for low-rate systems where very low outage probabilities are the main goal. As the rate is increased, direct transmission becomes more and more attractive.